\documentclass[conference]{IEEEtran}
\IEEEoverridecommandlockouts
\usepackage[utf8]{inputenc}
\usepackage{cite}
\usepackage{amsmath,amssymb,amsfonts}
\usepackage{booktabs} 
\usepackage{multirow}
\usepackage{subfigure}
\usepackage{tabulary}
\usepackage{graphicx}
\usepackage{enumitem}
\usepackage{xcolor}

\usepackage[disable, prependcaption, textsize=footnotesize, linecolor=red, backgroundcolor=red!25, bordercolor=red, textwidth=1.15cm]{todonotes}

\newcommand{\change}[1]{\todo[linecolor=orange,backgroundcolor=orange!25,bordercolor=orange]{#1}}

\newcommand{\ra}[1]{\renewcommand{\arraystretch}{#1}}
\newcommand{\subscript}[2]{$#1 _ #2$}

\def\BibTeX{{\rm B\kern-.05em{\sc i\kern-.025em b}\kern-.08em
    T\kern-.1667em\lower.7ex\hbox{E}\kern-.125emX}}
\begin{document}

\title{Measuring the engagement level in encrypted group conversations by using temporal networks}

\author{\IEEEauthorblockN{Mosh\'e Cotacallapa}
\IEEEauthorblockA{\textit{National Institute for Space Research} \\
S\~ao Jos\'e dos Campos, SP, Brazil \\
moshe@usp.br}
\and
\IEEEauthorblockN{Lilian Berton}
\IEEEauthorblockA{\textit{Federal University of S\~ao Paulo} \\
S\~ao Jos\'e dos Campos, SP, Brazil \\
lberton@unifesp.br}
\and
\IEEEauthorblockN{Leonardo N. Ferreira}
\IEEEauthorblockA{\textit{National Institute for Space Research} \\
S\~ao Jos\'e dos Campos, SP, Brazil \\
leonardo.ferreira@inpe.br}
\and
\IEEEauthorblockN{Marcos G. Quiles}
\IEEEauthorblockA{\textit{Federal University of S\~ao Paulo} \\
S\~ao Jos\'e dos Campos, SP, Brazil \\
quiles@unifesp.br}
\and
\IEEEauthorblockN{Liang Zhao}
\IEEEauthorblockA{\textit{University of S\~ao Paulo} \\
Ribeir\~ao Preto, SP, Brazil \\
zhao@usp.br}
\and
\IEEEauthorblockN{Elbert E. N. Macau}
\IEEEauthorblockA{\textit{Federal University of S\~ao Paulo} \\
S\~ao Jos\'e dos Campos, SP, Brazil \\
elbert.macau@inpe.br}
\and
\IEEEauthorblockN{Didier A. Vega-Oliveros}
\IEEEauthorblockA{\textit{University of S\~ao Paulo} \\
Ribeir\~ao Preto, SP, Brazil \\
davo@icmc.usp.br}
}

\maketitle

\begin{abstract}

Chat groups are well-known for their capacity to promote viral political and marketing campaigns, spread fake news, and create rallies by hundreds of thousands on the streets. Also, with the increasing public awareness regarding privacy and surveillance, many platforms have started to deploy end-to-end encrypted protocols. In this context, the group's conversations are not accessible in plain text or readable format by third-party organizations or even the platform owner. Then, the main challenge that emerges is related to getting insights from users' activity of those groups, but without accessing the messages. Previous approaches evaluated the user engagement by assessing user's activity, however, on limited conditions where the data is encrypted, they cannot be applied. In this work, we present a framework for measuring the level of engagement of group conversations and users, without reading the messages. Our framework creates an ensemble of interaction networks that represent the temporal evolution of the conversation, then, we apply the proposed  \emph{Engagement Index} (EI) for each interval of conversations to asses users' participation. Our results in five datasets from real-world WhatsApp Groups indicate that, based on the EI, it is possible to identify the most engaged users within a time interval, create rankings and group users according to their engagement and monitor their performance over time.
\end{abstract}


\section{Introduction}

Communication tools like e-mail and discussion forums are very common since the beginning of the Internet \cite{BICKART2001,Chatop2011}. 
Nowadays, Online Social Networks (OSN) and Messaging Apps (MA) added features to boost interactions and engagement between their users through group chats. According to the Global Digital Report 2019 \cite{Digital-2019}, the number of social media users worldwide in 2019 is 3.484 billion, increasing 9\% year-on-year. A compilation of the most popular social networks worldwide shows that Facebook holds the majority over 2 billion active users, followed by YouTube and WhatsApp \cite{Statista-2019}. 

In many countries, citizens have adopted messaging Apps, like WhatsApp, WeChat, Telegram, Viber, Line, etc., as the preferred medium for communication with their family, friends, coworkers, or clients. These platforms allow their users to create groups (chat rooms), in which massive and viral communication, about topics from religion to sports and politics, occurs between like-minded people. For example, WhatsApp Groups (WGs) is one of the main arenas for intense political or marketing campaigns, self-organized movements, among other activities in many countries. In India, for instance, the two major political parties claimed to have more than 20 thousand WGs that allowed to mobilize millions of sympathizers~\cite{bonis_2018}. The parties also had thousands of ``WhatsApp warriors'' broadcasting biased post in the groups, inflammatory political content, or fake news. In Brazil, in 2018 an audio with fake information about a pandemic with mortal victims was shared among several WGs, producing collective paranoia and chaos in public health services ~\cite{gisele_2018}.

Due to privacy issues in many social and messaging platforms, a huge quantity of private conversations and personal information were used for monitoring citizens, political campaigns, targeted ads, and many other initiatives from third-party companies and governments~\cite{chen_2018}.
In recent years, to improve security and privacy in communication, many of the MA have implemented end-to-end encrypted communication between users. However, the privacy and security offered by these systems have also been used for illegal activities or problematic behaviors, like the spread of rumors, spams, fake news, and the influence of the public opinion for arbitrary goals~\cite{gisele_2018,bonis_2018}. Although it is possible to encrypt the messages, several metadata attributes are available to capture. Then, with the appropriate tools is possible to infer the activities that a user have within a platform ~\cite{Park2016}.

Understanding users behavior and engagement on these platforms is quite important in many scenarios: a) For group owners or moderators, the relevance is related to have an efficient and quantitative way to measure the engagement of the group members, in order to know whom to reward, promote or remove from the group. Some solutions in this direction have been developed, but limited to moderate the group members and to show basic metrics like the quantity of messages or active members, without considering the engagement in terms of interaction with the whole group. b) It's well known that most of these platforms are free to use an their business model is based on a large advertising structure. Then, based on the engagement that a user have in a group (related to a topic), could be possible to target better or more interesting ads. c) Moreover, national security organizations could monitor dangerous or criminal groups, based only on their message activities, without requiring to read the messages content. To deal with this challenge, prior works have analyzed macroscopic features like session statistics \cite{Althoff_2015} or activity frequency \cite{Yang_2018}, return rate prediction \cite{Kapoor_2014}, or user retention \cite{Kawale_2009}. Few works used temporal information \cite{Holme2012,Gao2020,Li2017} that is natural in this kind of scenarios \cite{Liu_2019}. Most of previous works analyzed the user interaction or engagement based on platform features, like the number of comments, shares, ink-strokes, among others \cite{Chatop2011}. However, they are limited analyzing the engagement between the users in encrypted scenarios. 

Based on the construction of temporal interaction networks of the conversation, we introduce the \emph{Engagement Index} (\emph{EI}), an alternative approach for measuring the level of engagement in groups without breaking privacy. Experimental results show that the proposed index is capable of quantifying the engagement of individuals over time. Besides, the temporal snapshots of the conversation can be clustered in categories according to the z-score of the  \emph{EI} networks, and therefore, mining which users are more involved in each conversation category.  Moreover, a case study on WGs was performed and allowed to monitor the users' behavior during the presidential election period in Brazil, by using the time-series of \emph{EI} centrality.

Our main contributions are threefold: 1) A flexible framework for measuring the engagement of conversations and users, without considering the content of messages; 2) A classification of chats and ranking of users according to their engagement values; 3) A novel metric for analyzing and mining users' behavior in chat group conversations.

The remaining of the work is organized as follow: Section \ref{related_work} presents some related work on user engagement and OSN analysis. Section \ref{research_problem} brings the research questions encompassed in this work. Section \ref{matherials_methods} shows the materials and methods employed, the temporal network construction, the proposed \emph{Engagement Index} (\emph{EI}) and the dataset used. Section \ref{results} has the results and discussion of the proposed framework and Section \ref{case_study} a case study on WGs to detect different user's behavior. Finally, Section \ref{conclusion} presents the final remarks and future work.

\section{Related work} \label{related_work}

Social network sites and MA raised great attention since their beginning in the mid-2000s. Within this context, the engagement has been related to several aspects like \cite{McCay-Peet2016}: 1) self-representation; 2) participation (liking, retweeting, etc); 3) use purpose (information, social activity); 4) positive experience that maintains user's engagement; 5) social context. Also, many studies used different techniques to capture information from the sessions, such as the time spent, or users' interactions~\cite{Chatop2011}.

Several measurements of social network engagement have been proposed, such as Facebook Intensity (FBI) that examined the association between Facebook engagement and social capital \cite{Ellison_2007}. In \cite{SIGERSON_2018}, the authors presented a systematic review of some measurements for social network engagement. The study pointed out that most scales were limited by their sample homogeneity and focused entirely on Facebook, which limits knowledge advancement in research on engagement with social networks as a whole.

Furthermore, in \cite{Thomas_2018}, the authors used a Temporal Convolutional Network to understand the intensity of engagement of students attending video material from Massive Open Online Courses (MOOCs). Similarly, a temporal evolving action graph was proposed by \cite{Liu_2019} to analyze mobile social apps characteristics in terms of informing future user engagement.

Given that group conversations are social interactions between group members, some studies developed methods to analyze the social behavior based on network science and computational approaches. For example, the discovery of roles and topic suggestion from the analysis of content messages and their direction (author - recipient) \cite{McCallum2007}; the detection of spammers in social networks by analyzing the behavior and content sent through messages \cite{Zheng2015}; and the analysis of friendship in social networks based on the sentiment analysis of messages exchanged between users \cite{Thelwall2010}.

\section{Research problem} \label{research_problem}

As previously mentioned, there is an increasing interest and concern on implementing new solutions that provide an optimal trade-off between privacy and information. The larger number of approaches that study users' engagement and behavior are based on content analysis and friendship network, which are not suitable in this scenario.

Recalling that in MA we do not have access to any personal information (contacts or attributes) or messages (in plain text) since they are encrypted, the critical point to tackle here is: \textit{How to measure users' engagement on group conversations without reading the content of their messages?}. We break down this point into the following research points: 
\begin{itemize}
	\item How to appropriately represent and construct the network of user interactions over time? Moreover, how to characterize the networks?
	\item Can we classify users according to the pattern of interaction and establish their levels of engagement within the group? 	
	\item From the behavioral pattern of users, could it be possible to identify the following profiles?: ($1$) The most engaged or ($2$) influential users; and ($3$) to find similar patterns of behavior?
\end{itemize}

\begin{figure}[t]
\centering	
\includegraphics[width= 0.51\textwidth]{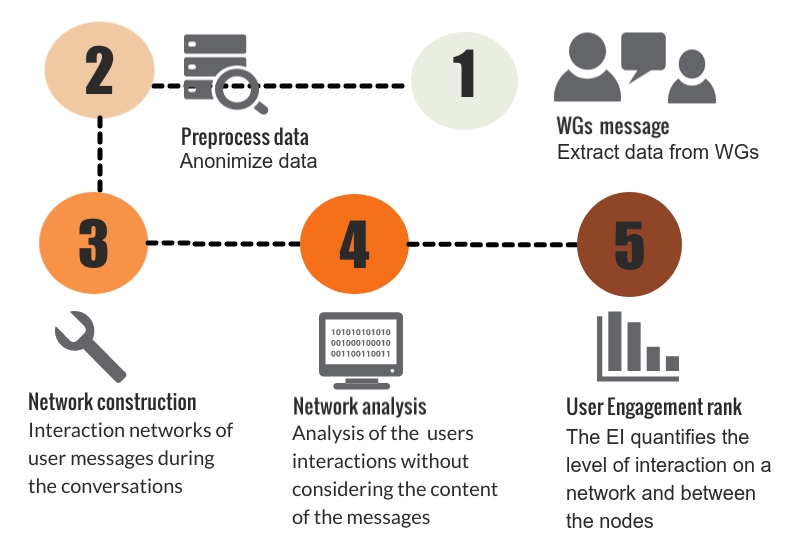}
\caption{Scheme of the data collection process from the WhatsApp groups: From the exporting tool of the conversations until the construction of the temporal interaction networks and network analyses.}
\label{fig:dataflow}
\end{figure}

This work is a first step in this direction, where we aim to analyze the users' behavior when sending messages in AM, from a network sciences perspective. Our approach consists in constructing an \emph{Interaction Network} from the messages sent by users in a specific time interval ($\Delta{t}$), see Figure~\ref{fig:dataflow}. Then, we obtain an ensemble of interaction networks that represent the temporal evolution of the user's activity into the WG. We employ temporal networks as a natural path for analyzing this ensemble of networks~\cite{Holme2012,Li2017,Gao2020}, where each layer represents a snapshot of user's activity in a particular $\Delta{t}$, and the inter-layer  connections represent the temporal evolution of users. Therefore, the temporal approach allows the mapping of behavioral patterns of social interaction in local, intermediate and global scales of the evolutionary process. To the best of our knowledge, this is the first approach that tackle the mining of users’ interaction and engagement in encrypted group conversations by employing temporal networks.

\section{Material and methods} \label{matherials_methods}

In network sciences, data can be represented by a static graph\footnote{The terms ``network'' and ``graph'' share the same definition and are interchangeable in this document} $G = (V, E, W)$,  where $V = \{v_1, v_2, \ldots,$ $v_n\}$ is the set of $n$ nodes, the set of $m$ links $E = \{e_1, e_2, \ldots, e_m\}$ that connects the nodes and the set $W$ of $m$ weights, one for each edge. However, when consolidating the temporal information in a static network, we loose part of the dynamics. In this sense, it is not possible to evaluate the performance and role of the nodes, nor understanding the interaction patterns into the network. As alternative, a temporal network ($\mathcal{G}$) can be represented as an ordered sequence of network observations at different time-steps or intervals~\cite{Gao2020}, i.e., $\mathcal{G} = \{G_0, G_1, \ldots, G_{l}\}$ with $l$ the number of layers or snapshots. In other words, for temporal networks, we have a long sequence of symmetric pairwise interactions representing observations over time. This dynamical network contains not only the set of similarity or relation links between nodes but also information on how the connection behavior evolves. 

Nevertheless, the data are not always ``naturally'' represented by a graph, but rather by events or time series. Thus, for applying the network techniques, a network must be constructed. One approach for reconstructing data into networks is upon the process of linking nodes according to the co-occurrence of events in a chronological fashion~\cite{Ferreira2020}, like in earth
sciences~\cite{Vega-Oliveros2019}, ecology \cite{Barberan2011}, text mining \cite{Vega-Oliveros2019a}, among other domains.

Here, we represent the behavior of users sending information in the group by interaction networks. These networks are the description of the co-occurrence of users' messages during the conversations, given the constraint that the content is encrypted. In this way, we analyze the roles of users concerning their interactions rather than the messages information.

\begin{figure}[t]
\centering	
\includegraphics[width= 0.4\textwidth]{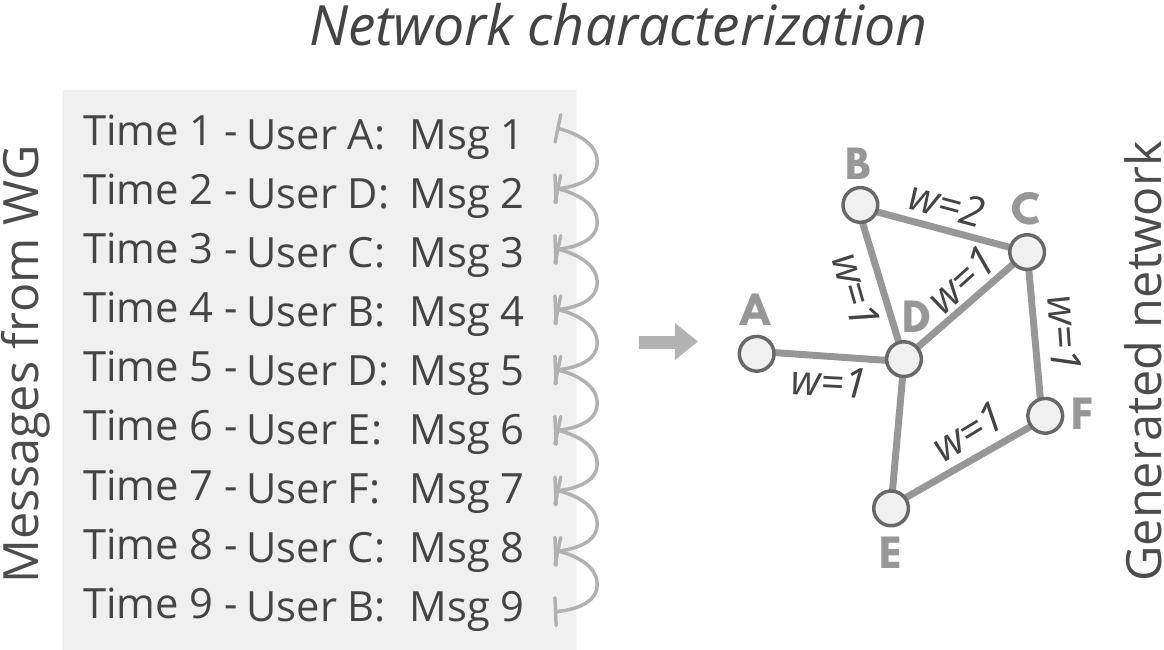}
\caption{Example of interaction network construction considering the co-occurrence of messages sent among users. Multi-edges are represented as weights between the nodes.}
\label{fig:network-model}
\end{figure}

\begin{figure*}[t!]
\centering	
\includegraphics[width= 0.85\textwidth]{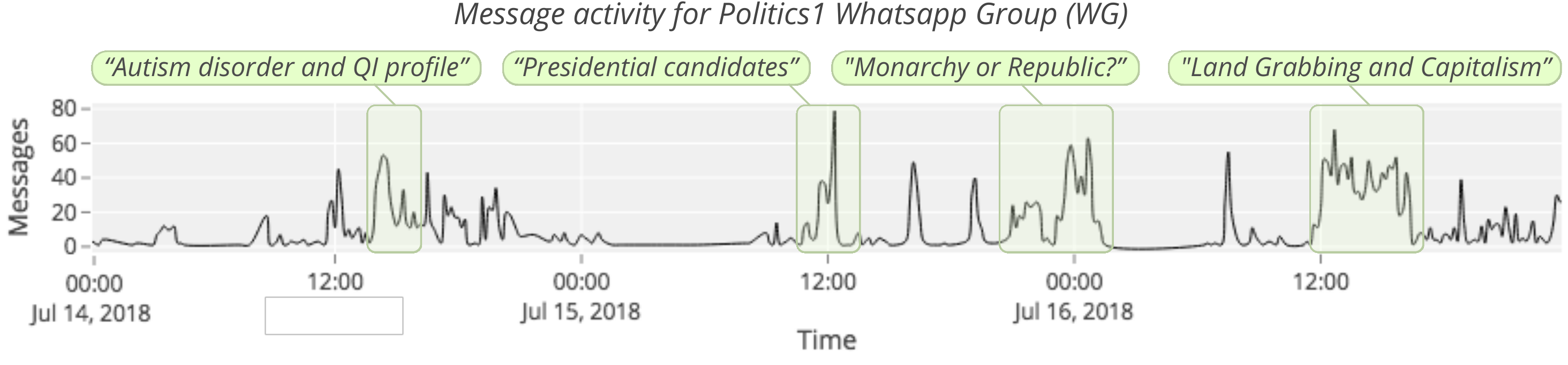}
\caption{Message activity that verifies the number of messages in a  time-interval $\Delta{t} = 10 min$.}
\label{fig:timeline}
\end{figure*}

\begin{figure*}[t!]
\centering	
\includegraphics[width= 0.85\textwidth]{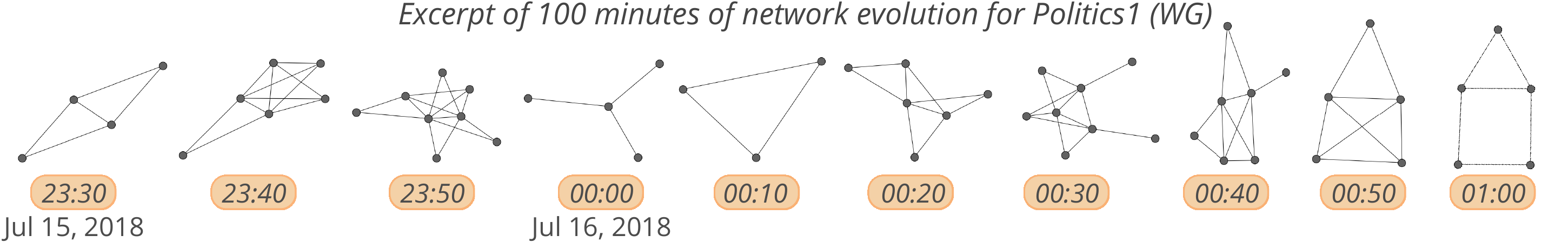}
\caption{A sequence of networks generated in each time interval.}
\label{fig:excerpts}
\end{figure*}

\subsection{Modeling messages behavior via interaction networks}

The users' patterns of interactions play a fundamental role to define their ability to propagate information or influence in their group. However, we need first to reconstruct the data into a network of interaction. Since we are working with encrypted group conversations, only basic metadata is available for our framework: the message time stamp and an identifier of the sender. We anonymize the user identification (step 2 of Fig.~\ref{fig:dataflow}) and construct the network (step 3 of Fig.~\ref{fig:dataflow}): each sender is represented by a node; following the chronological order, a link is created between two senders if they send messages  one after the other. To measure the network behavior, we divide the whole time period in time intervals ($\Delta{t}$), to obtain an ensemble of interaction networks that represent the temporal evolution of the conversation group. The decision to have a fixed interval of time is based in two reasons: 1) To give equal conditions to create the temporal networks, because it is expected that the longer the time, the more interactions are captured by the network. 2) To monitor the evolution of the engagement over time, for each user and conversation. 

After that, we simplify each network by removing self-loops since our purpose is to study dialogues and not monologues. We assume that the messages are broadcast to all the members and not directed to a specific individual. Therefore, the links are undirected and weighted. This modeling allows the mapping of behavioral patterns of social interaction in local, intermediate, and global scales as an evolutionary process.

A hypothetical example of the network construction process is illustrated in Figure~\ref{fig:network-model}.
Node A corresponds to Phone A (User A), node D to Phone D (User D) and so on. Each time a node sends a message, it is connected to the node of the previous message. For instance, node D sent a message after A at time 2, and thus, in the interaction network, they are connected. We generate weighted and undirected networks avoiding self-loop connections. Here, the multi-edges are represented as the sum of edges between two nodes and used as the weights of the connections. This graph representation, which is a snapshot of the message activity of users, differs when compared to only considering the number of messages. For example, if a user sends 30 messages in a row, s/he is merely interacting alone, with a low group engagement in the conversation. For this reason, we define that is necessary at least two users interacting to be considered as a conversation.

We select as an example the message activity in a WhatsApp Group (Politics1), collected over three days, with the count of messages in intervals of $\Delta{t} = 10$ minutes. The reason to choose $10$ minutes is based on several experiments made in areas like interpersonal communications and analysis of behavior, that consider this time enough to have a conversation or discussions \cite{Azrin1961, Fant2015}. As we collected those messages in plain text, for illustrative purposes, we tagged some moments with the particular topic discussed at the moment, as we can see in Figure~\ref{fig:timeline}. However, it is important to highlight that we do not use the content messages on our method. 

Considering that any discussion has different stages over time, by using the network characterization, we generate a sequence of networks, each one representing time slices of interactions. The generated networks present several topology, indicating that the interaction networks have no trivial or regular connections. In this way, we can extract some patterns from the topological information of the ensemble. For instance, in the previous example the network at 23:30 of July 15 has $4$ nodes as the midnight network of the same day. However, the structures are different, as shown in Figure~\ref{fig:excerpts}. For this reason, we present a new measure, called \emph{Engagement Index} (EI), which seeks to quantify the engagement regarding users' interaction on the network, presented as follow. 

\subsection{The Engagement index (EI)}

Here, we consider that the engagement is high in a conversation group if a great number of users participate homogeneously in the conversation. Which means that the topic is interesting enough to get the attention and participation of the members. For this purpose, we evaluate not only the sequence of messages, but also, how many users were interacting and how equally was that interaction.

Formally, we define the EI in terms of the equality and intensity of the network interactions, i.e.,

\begin{align}
\label{eq:enga} \hbox{\emph{Engagement Index}} (G) & =  \hbox{\emph{Equality}}(G) * \hbox{\emph{Intensity}} (G) \\
\label{eq:equaly}  \hbox{\emph{Equality}} (G) & = 1 - \hbox{\emph{Gini}}(W) \\ 
\label{eq:intense}  \hbox{\emph{Intensity}} (G) & = \log_2 (n*\frac{1}{2}\sum_{i}^{m}w_i)
\end{align} where the \emph{Engagement} is the product between the \emph{Intensity} and \emph{Equality} of users' interaction on the network. 

The \emph{Equality} is the complement of the \emph{Gini} coefficient, originally proposed for measuring the level of inequality in the incoming of a population~\cite{Gastwirth72}. The \emph{Gini} values vary from $0$ (full equality) to $1$ (total inequality). Therefore, in Eq.~\ref{eq:equaly} we are interested in measuring how equally was the message interactions (weighted links) among the participants. 

In Eq.~\ref{eq:intense}, we have that all weights $w_i\in W$ are positive integers greater than zero. With the \emph{Intensity} we measure how intense was the conversation (network) in terms of the number of participants (nodes) and the total user-to-user messages (links). The \emph{Intensity} is equal to $1$ when the network has at least two nodes interacting once, i.e., with an average degree equal to $1$. This is the reason for the use of $\log_2$ in the measure. The \emph{Intensity} is equal to zero in the case of a network with a single node, which is not considered as a group conversation.

The \emph{EI} is, therefore, the combination of high message activity between a large number of individuals but with a more equally distributed participation. For example, conference talks or broadcasting messages, where only one source is expressing its ideas, is not ideally considered as a conversation with high engagement. We present some illustrative examples of interaction networks and their \emph{EI} values in Table~\ref{tab:toynetworks}.
\begin{table}[!t]
    \caption{Toy networks depicting different cases of applying the proposed \emph{Engagement} index. For illustrative purpose, we plot the networks showing multi-edges interactions.}
    \flushleft
     \begin{flushright}
		   \subfigure[]{\label{fig:toy1}\includegraphics[scale=0.8]{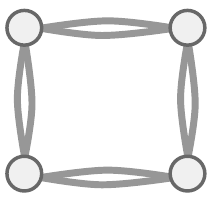}} \hspace{0.25cm}
           \subfigure[]{\label{fig:toy2}\includegraphics[scale=0.8]{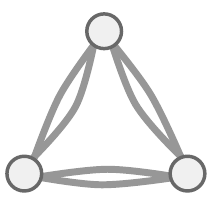}} \hspace{0.25cm}          \subfigure[]{\label{fig:toy3}\includegraphics[scale=0.8]{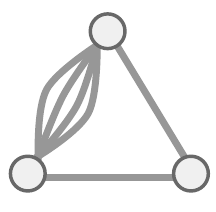}} \hspace{0.25cm}
           \subfigure[]{\label{fig:toy4}\includegraphics[scale=0.8]{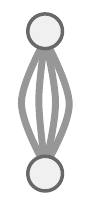}}
	\end{flushright}
       \vspace{-0.3cm}
       \ra{1.3}
       \begin{tabular}{@{}lc cc cc cc c@{} }
       \cmidrule{2-9}
       $n$ && 4 &\phantom{abcdefghi}& 3 &\phantom{abcdefg}& 3 &\phantom{abcd}& 2 \\
       $m$ && 8 && 6 && 6 && 4 \\
       \emph{Intensity} && 5 && 4.17 && 4.17 && 3 \\
       \emph{Equality} && 1 && 1 && 0.83 && 1 \\
       \emph{Engagement} && 5 && 4.17 && 3.47 && 3 \\
       \bottomrule
       \end{tabular}
    \label{tab:toynetworks}
\end{table}
First, we have in columns \subref{fig:toy1}, \subref{fig:toy2}, and \subref{fig:toy4} three examples of network with homogeneous distribution of links between the nodes. Therefore, the \emph{Equality} of the networks is the highest. However, the network in \subref{fig:toy1} reach a higher \emph{EI} due to a larger number of participants than networks \subref{fig:toy2} and \subref{fig:toy4}. On the other hand, the network in \subref{fig:toy3} has the same number of nodes, links, and \emph{Intensity} than network \subref{fig:toy2}, but most of the interactions are concentrated in only two nodes, producing a drop in the \emph{Equality} of the network \subref{fig:toy3}. For this reason, network \subref{fig:toy3} reaches a lower \emph{Engagement} value than network \subref{fig:toy2}. The computational cost of the \emph{EI} is dominated by the order of calculating Eq.~\ref{eq:equaly} ($O(n*log(n))$) or Eq.~\ref{eq:intense} ($O(m)$), i.e., $O(\emph{EI}) = \max[O(\emph{Gini}), \ O(\emph{Intensity})] \equiv \ O(m)$.

Regarding the \emph{EI} for the nodes, we define the \emph{Engagement} centrality as the \emph{EI} value of the network proportional to the participant interaction with respect of all the nodes, i.e.,
\begin{align}
    \hbox{\emph{EI}}(G, v_i) & = \frac{n *  w_i * \hbox{\emph{EI}}(G)}{2*m}
    \label{eq:engaCentrality}
\end{align} where $w_i$ is the weighted degree of node $v_i$. This way, each node contributes proportionally to the network \emph{Engagement} according to their number of interactions ($wk_i$), which means, the \emph{EI} of the network is the average of the \emph{EI} of the nodes. 

\subsection{Data collection}

We focused on data from Brazilian WG that address different topics. For this purpose, we extracted the conversation history (for at least 60 days) along 2018, from five groups, related to: (1) Theology (one group), due to the wide variety of religious society composed by thousands of denominations; (2,3) Politics (two groups), given the corruption scandals, and the presidential elections that occurred in October of that year; (4) English (one group), a WG of enthusiastic learners of English as a second language; and (5) Vegetarianism (one group), a group promoting a vegetarian lifestyle. The anonymized datasets are available to download in \cite{co_2019}. 
 
\begin{table}[h]
\caption{Characteristics of the five datasets.}
\centering
\begin{tabular}{@{}l llll@{}}
\toprule
Groups     & \# Messages & \# Users & From       & To         \\ \midrule
Politics1  & 79082       & 489      & 2018-09-02 & 2018-11-20 \\
Politics2  & 78319       & 628      & 2018-08-09 & 2018-11-20 \\
Vegetarian & 10593       & 120      & 2018-08-15 & 2018-11-20 \\
English    & 11325       & 218      & 2018-09-07 & 2018-11-20 \\
Theology   & 70213       & 304      & 2018-07-24 & 2018-11-20
\end{tabular}
\label{tab:dataset-metrics}
\end{table}

Table~\ref{tab:dataset-metrics} shows the quantity of messages and active users over time for the five WGs. The data gathering process started with five mobile devices, each one with a WhatsApp account and a membership in the selected topics of WGs. As shown in Figure~\ref{fig:dataflow}, the whole process is divided into the following steps:
First, we export the WG data using the ``Export conversation'' option located inside the group chat settings. This feature allowed us to collect and save all the messages in plain text. Then, in step two, we preprocess the messages by cleaning and anonymizing the data. At the end of this step, we only save a log of the messages for each WG, which contains two columns: user ID (who sends the message) and timestamp (the time the message was sent), then we save it in a database to be easily accessed. In step three, we start the construction of the temporal interaction networks, which are the input for our framework. Each log entry is divided into fixed intervals of time, and for each interval, an interaction network is built. Therefore, for each WG we have at least 60 days of continuous data. 
In total, by considering slices of 10 minutes, we have $16732$ conversations (characterized as undirected networks) for the five WGs, after removing self-loops but preserving multiple edges between nodes. In step four, the generated networks are analyzed and finally, in step five, the engagement characterization is performed for each interaction network.

\section{Experimental results} \label{results}

\begin{figure}[t]
\centering	
\subfigure[]{\label{fig:x-intensity-engagement}\includegraphics[width= 0.35\textwidth]{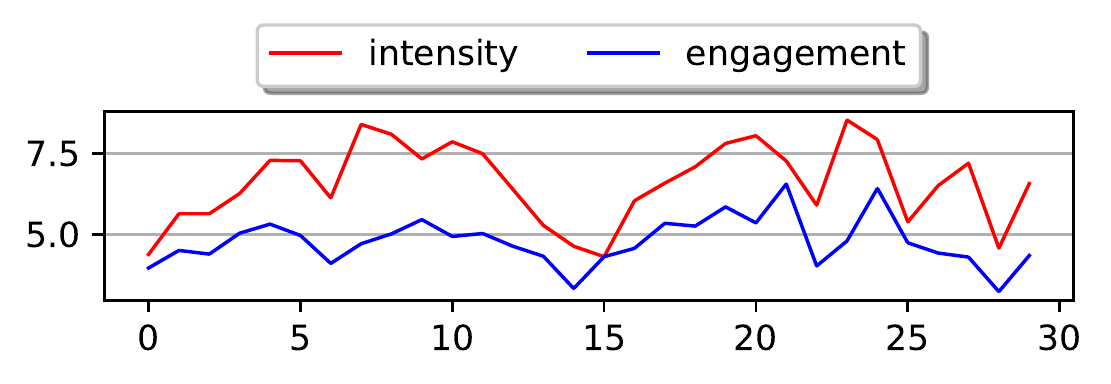}}\vspace{-0.2cm}
\subfigure[]{\label{fig:x-equality}\includegraphics[width= 0.35\textwidth]{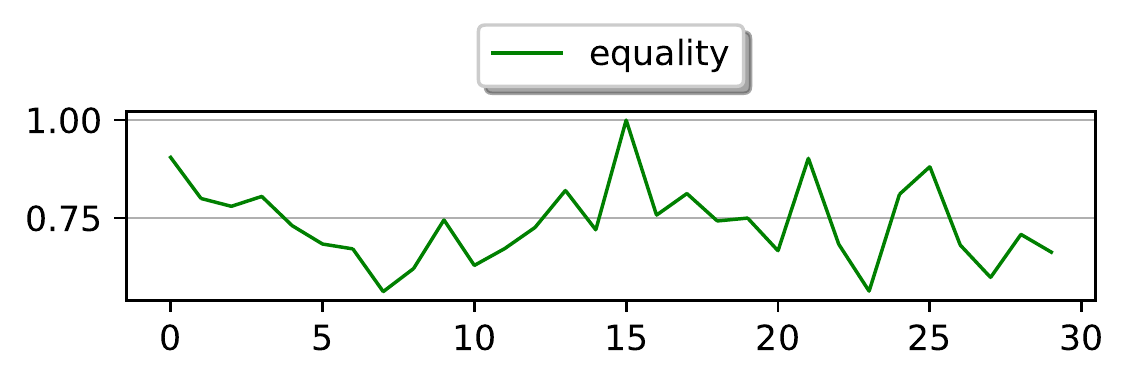}}
\caption{\small (Color online) Sample of 30 networks from the Politics1 group: (a) the \emph{Engagement} and \emph{Intensity} values of the networks; and (b), the \emph{Equality} measurements}
\label{fig:equality-engagement-intensity}
\end{figure}

We applied our proposed framework to construct the temporal networks of interactions for each WG described in Tab. \ref{tab:dataset-metrics}. Specifically, we are interested in understanding the role of the users in terms of the network \emph{Engagement} over time. We first calculate the \emph{EI} for all temporal networks from the WGs. We show a sample of the \textit{intensity, engagement and equality} values for 30 temporal networks constructed using the Politics1 data set (Figure~\ref{fig:equality-engagement-intensity}). For the first 12 networks, the \textit{EI} values remains stable around 4 and 5, while the \textit{Intensity} and \textit{Equality} values have significant changes. On the other hand, for the last 18 networks, all the variables changed considerably. \change{This means that...}

\begin{figure}[h!]
\centering	
\subfigure[]{\label{fig:histpolitics1}\includegraphics[width= 0.23\textwidth]{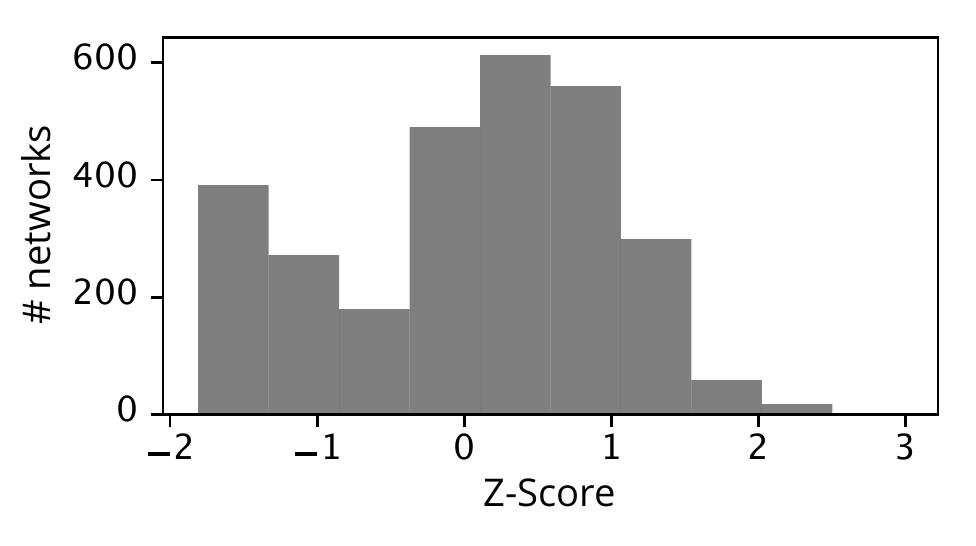}}
\subfigure[]{\label{fig:histpolitics2}\includegraphics[width= 0.23\textwidth]{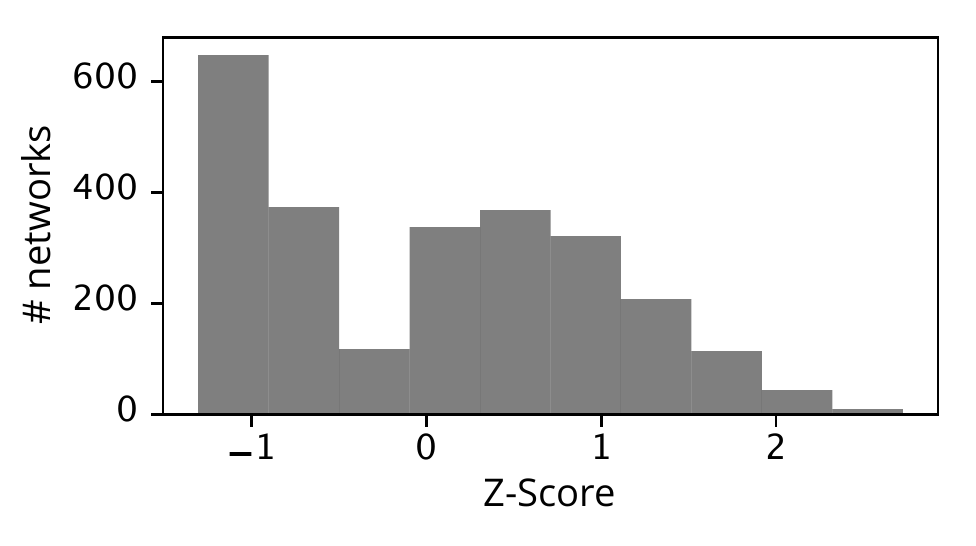}}
\subfigure[]{\label{fig:histvegetarian}\includegraphics[width= 0.23\textwidth]{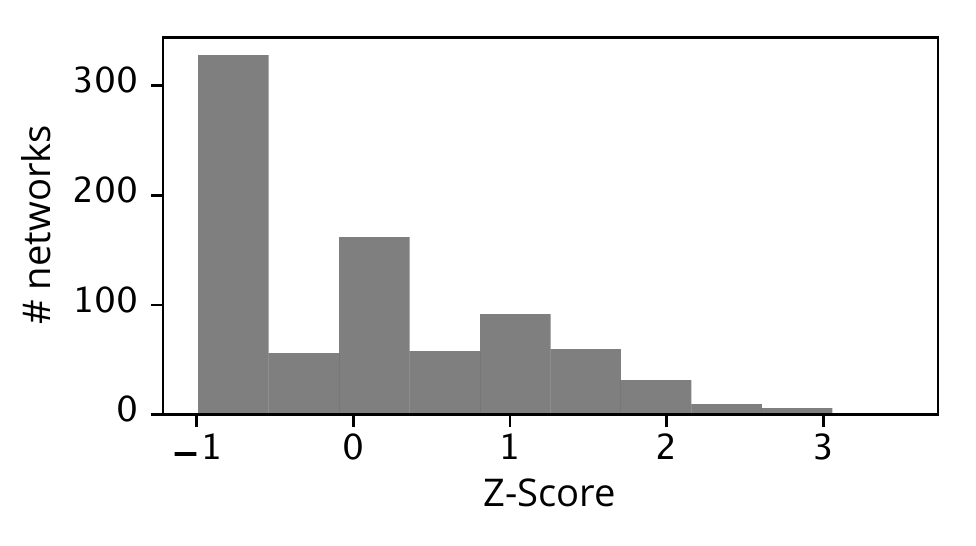}}
\subfigure[]{\label{fig:histenglish}\includegraphics[width= 0.23\textwidth]{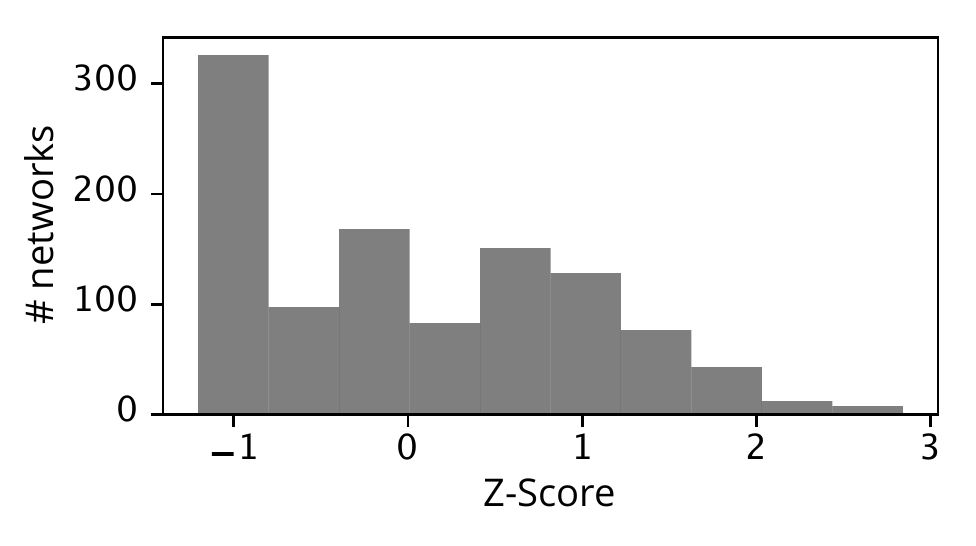}}
\subfigure[]{\label{fig:histheology}\includegraphics[width= 0.23\textwidth]{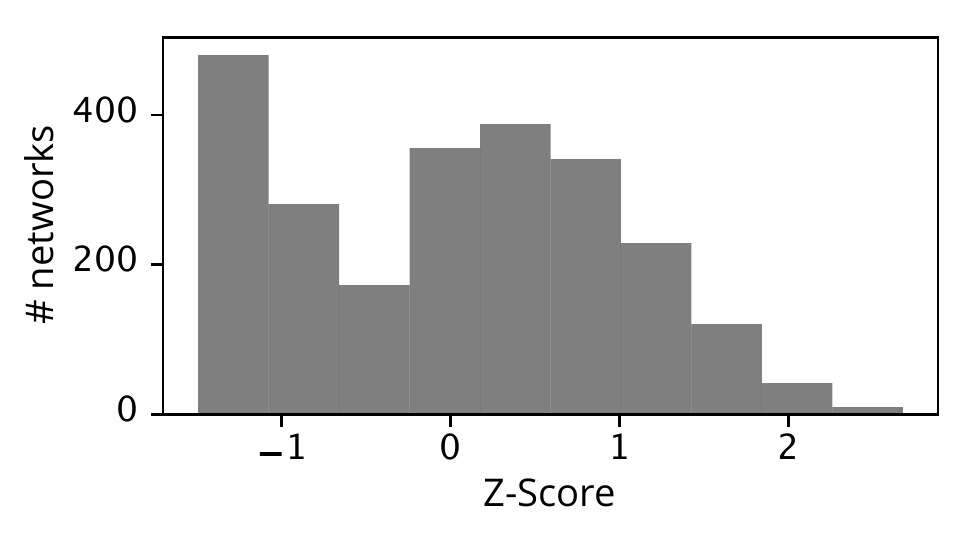}}
\subfigure[]{\label{fig:histall}\includegraphics[width= 0.23\textwidth]{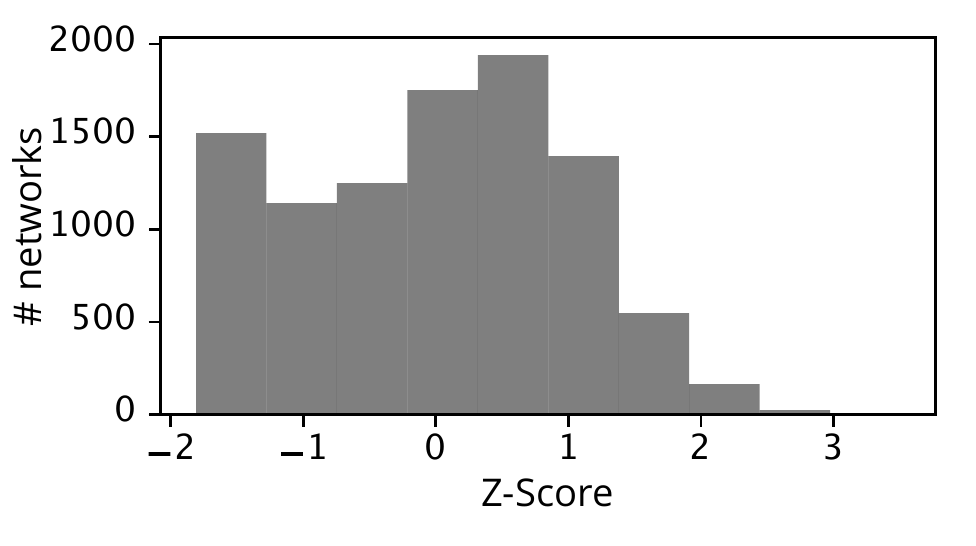}}
\caption{ Histogram of the z-score of \emph{EI} values for the collected WGs: \subref{fig:histpolitics1} Politics1 group; \subref{fig:histpolitics2} Politics2 group; \subref{fig:histvegetarian} the Vegetarian group; \subref{fig:histenglish} the English group; \subref{fig:histheology} the Theology group; and \subref{fig:histall} the aggregation of all the groups together.}
\label{fig:histWGs}
\end{figure}

To compare the similarities and differences across the groups, we calculate the z-score values of the networks by WGs, defined as: 

\begin{equation}
\hbox{z-score}(G_j) = \frac{\hbox{\emph{EI}}(G_j) - \mu[{\hbox{EI}(\mathcal{G})}]}{ {\sigma[{\hbox{ EI}(\mathcal{G})}]}}, \ G_j \in \mathcal{G},
\end{equation} 

\noindent where $\mu[{\hbox{ EI}(\mathcal{G})}]$  and $\sigma[{\hbox{ EI}(\mathcal{G})}]$ are respectively the mean and the standard deviation of the \emph{EI} values for all the networks. We present the histograms of the z-score values for each dataset (Figure~\ref{fig:histWGs}). We can observe that each WG has particularities in terms of the level of \emph{EI} over the networks. In Politics2, English, and Theology groups, a portion of the networks with \emph{EI} values have one standard deviation below the mean. Another fraction of networks have \emph{EI} values close to the mean, and some other networks have high engagement values, above one standard deviation of the mean. Given the before considerations, we classify the networks in three z-score categories: HIGH engagement networks, with z-scores values greater than or equal to $1$; MEDIUM engagement networks, with z-score values between $(-1, 1)$; and LOW engagement networks, with z-score values below or equal to $-1$.
  
The proposed \emph{EI} classification (\emph{EI}C) is crucial for understanding the dynamics of high, medium, or low engagement in group conversations. This network classification can also be extended to the nodes, in which the participants can have different roles depending on the message interaction they have in the network classes. Users that are more representative in LOWER engagement networks can be identified as initiators, claimers, or finishers of the discussion topics of the group. Opposite, representative users in the HIGHER \emph{EI} networks, can be seen as conciliators, or argumentative users expressing strong positions in the conversation.

\begin{figure}[b]
\centering	
\includegraphics[width= 0.45\textwidth]{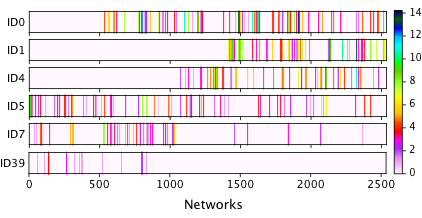}
\caption{ \emph{Engagement} values for six particular users from Politics2 WhatsApp Group for more than 2500 networks.}
\label{fig:six-users}
\end{figure}

\begin{table}[h]
\caption{Nodes ranking of the average \emph{EI} of the temporal networks from the Politics2 WG, following the \emph{EI} classification: the group of HIGHER z-score values; the group of MEDIUM z-score; networks in the group with LOWER z-score values; and the GLOBAL ranking among all the valid networks.}
\centering
\resizebox{0.48\textwidth}{!}{
\begin{tabular}{@{}lllllllllll@{}}
\toprule
\multicolumn{2}{c}{HIGH} && \multicolumn{2}{c}{MEDIUM} && \multicolumn{2}{c}{LOW} && \multicolumn{2}{c}{GLOBAL} \\
\cmidrule{1-2} \cmidrule{4-5} \cmidrule{7-8} \cmidrule{10-11}
 ID & Mean && ID & Mean && ID & Mean && ID & Mean \\
0 & 2.71944 && 0 & 0.78728 && 7 & 0.11282 && 0 & 0.95457 \\
1 & 2.10064 && 2 & 0.61945 && 5 & 0.11128 && 1 & 0.68665 \\
2 & 1.61163 && 5 & 0.55969 && 3 & 0.09273 && 2 & 0.65913 \\
4 & 1.54028 && 3 & 0.55510 && 6 & 0.06955 && 3 & 0.57078 \\
3 & 1.28391 && 1 & 0.52203 && 2 & 0.06182 && 4 & 0.55478 \\
6 & 1.13795 && 7 & 0.51909 && 39 & 0.05873 && 5 & 0.49139 \\
8 & 1.08223 && 4 & 0.46532 && 0 & 0.05409 && 6 & 0.47743 \\
5 & 0.81126 && 6 & 0.44704 && 20 & 0.05100 && 7 & 0.41087 \\
10 & 0.78318 && 9 & 0.26586 && 26 & 0.04482 && 8 & 0.34396 \\
11 & 0.72572 && 8 & 0.24997 && 50 & 0.04327 && 9 & 0.26929 \\ 
\bottomrule
\end{tabular}
}
\label{tab:eng-ranking}
\end{table}

After calculating the \emph{EI} centrality for the nodes in all the WG networks, we have the \emph{EI} centrality for six users from the Politics2 group (Figure~\ref{fig:six-users}).  Each bar represents the temporal \emph{Engagement} evolution of the user. We notice that this time-series characterize the message behavior of the users, showing the moments in which they interact with the others and their levels of engagement or relevant participation in the group.

We separate the networks into three groups according to the \emph{EI}C. Then, for each node in the group of networks, i.e., in the LOWER, MEDIUM, or HIGHER, we calculate the average \emph{EI} centrality in the group. Additionally, we calculate the average \emph{EI} centrality of the nodes considering all the networks, which we call as the GLOBAL group. As an example, the top 10 ranking \emph{EI} centrality nodes by \emph{EI}C are reported for the Politics2 WG (Table~\ref{tab:eng-ranking}). The IDs of the users are according to the GLOBAL ranking, which are the same in Figure~\ref{fig:six-users}. 

In the Table~\ref{tab:eng-ranking}, we have that the GLOBAL top ranked user is also the best ranked in the HIGHER and MEDIUM group. In this particular case, this user is the moderator/manager of the group, which is a very active participant. However, this is not the natural tendency for the other groups. Comparing the rankings in each classification group, we can observe some position differences between the users. The difference in \emph{Engagement} behavior is notable according to the ranking in the classes and the Figure~\ref{fig:six-users}: ID 5 has more regular participation during the discussion over time, but it is better ranked in MEDIUM and LOW \emph{EI} networks. Opposite, ID 0 and ID 1 have meaningful participation in GLOBAL and in HIGH \emph{EI} networks. However, user ID 1 tends to be less engaged in LOW or MEDIUM \emph{EI} networks. Users of ID 7 and ID 39 are better ranked in LOW \emph{EI} networks than in the others, indicating the tendency of interacting in low activity moments.

\section{Behavior comparison between groups} \label{case_study}

Here, we aim to identify the differences in the users' behavior on three WGs: Politics2, Vegetarian and Theology, specifically during the presidential elections that happened in Brazil in 2018. The following two intervals where considered:

\begin{enumerate}[label=(\subscript{P}{{\arabic*}})]  
    \item \textit{Pre-electoral}: From first of October to October 28 of 2018.
    \item \textit{Post-electoral}: After Brazilian presidential elections, from October 29 to November 21 of 2018
\end{enumerate}

To discover whether the presidential election has a different influence over the Politics group than the others, as a first step, we characterize both periods (\textit{Pre-electoral} and \textit{Post-electoral}) using the proposed framework of temporal interaction networks. The \emph{EI} values are calculated for all conversations (networks with $n > 1$). Then, we calculate the average \emph{EI} centrality for each user considering: (\textit{i}) the whole period ($P_1$ + $P_2$); and (\textit{ii}) the  periods $P_1$ and $P_2$ separated. In each case, we obtain three vectors of the average \emph{EI} centrality of the nodes, each one corresponding to $P_1, P_2, \hbox{ and } (P_1$ + $P_2)$. The vectors are ordered according to the sorting of the whole period vector, and each one is normalized by its highest \emph{EI} value. 

\begin{figure}[t]
\centering	
\includegraphics[width= 0.4\textwidth]{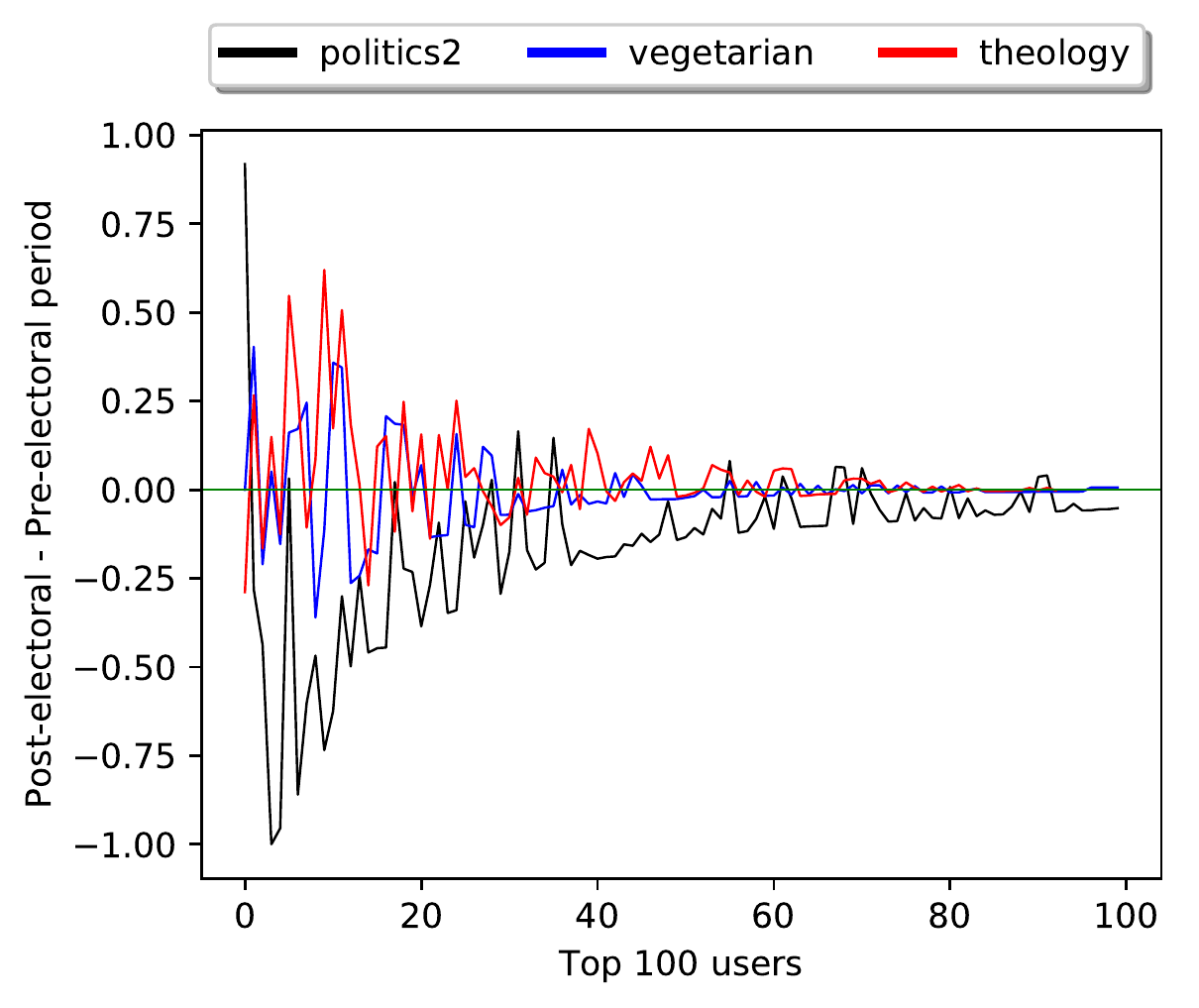}
\caption{ Normalized  difference of \emph{Engagement} values (\textit{Post-electoral} - \textit{Pre-electoral} period) for three WGs: Politics2, Vegetarian and Theology.}
\label{fig:ranking-differences}
\end{figure}

We compare the differences between $P_2$ and $P_1$ for all the users. This way, negative values mean that the \emph{EI} of the users before the presidential elections were higher than after the elections, and positive values otherwise. The top 100 users in the whole period and their differences values are in Figure~\ref{fig:ranking-differences}. Each line corresponds to one of the three WGs. Notably, the difference between $P_1$ and $P_2$ is more intense for Politics2 than the other two WGs. Some users reached more negative difference values. Five of the top members (within top 10) reached values lower than $-0.61$. The before indicates that these users, who had a high engagement before the elections, abruptly stop interacting after $P_1$. Clearly, this behavior is not repeated in the other two WGs groups (theology and vegetarian).

\begin{figure}[h]
\centering	
\includegraphics[width= 0.4\textwidth]{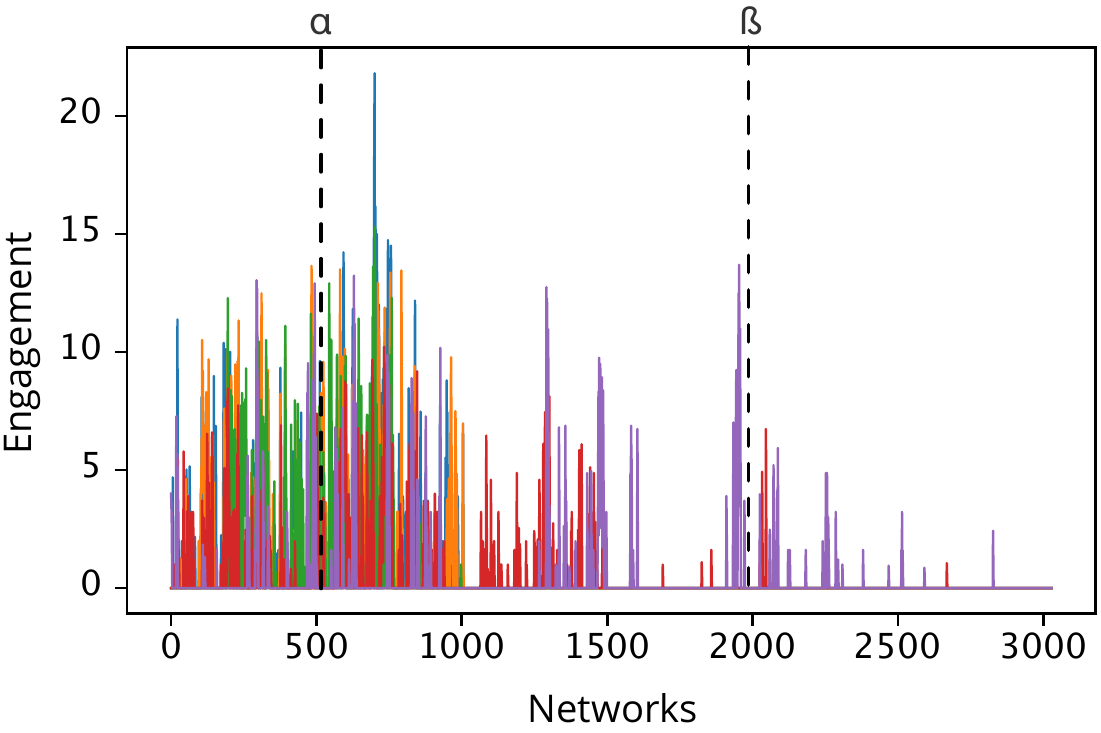}
\caption{ \emph{Engagement} values for top five individuals with most negative normalized \textit{EI} difference (\textit{Post-electoral} - \textit{Pre-electoral}) $<-0.61$ in Politics2 WG. ($\alpha$) First and ($\beta$) second round for presidential elections in Brazil (2018).}
\label{fig:anomalous-users}
\end{figure}

To visualize how those five users were engaged over time, we show their \textit{EI} values, where each network represents a conversation (Figure~\ref{fig:anomalous-users}). Note that although these users lose engagement, some of them are still active at some points after $P_2$ (point $\beta$ in the Figure). We highlight that the proposed index is flexible and suitable for different time windows. For example, we can compare the gain or loss of engagement in weekly/monthly intervals, or even in real-time, and find the members with engagement anomalies over time.

\section{Conclusion} \label{conclusion}

In this work, we propose a framework for analyzing users' behavior in encrypted group message applications. We employ temporal interaction networks to represent the user-user message interaction over time. Given the encryption constraint, we introduce the \emph{Engagement Index} (\emph{EI}) to measure the level of participation of the users according to their messages behavior. We tested the framework with data collected from five groups of WhatsApp and a variety of topics.

By mining this data with the \emph{EI} measure, it is possible to rank engaged users and groups. In our understanding, this project contributes to opening a new path to identify interaction users' patterns in encrypted messages groups. Also, the \emph{EI} could be beneficial for moderators to quickly check the group engagement, rank users according to their performance, and identify anomalous contributors.

As future works, several analyzes can be performed following or extending the proposed framework, applying not only in WGs but also in other similar platforms and environments, like Telegram Groups, Forums, or Live stream chats. Also, other potential uses are the detection of temporal patterns from the interactions, correlating the engagement values between users, and the characterization of group topics according to the users' engagement over time. Furthermore, the interaction networks can be analyzed considering motif-based patterns~\cite{Leskovec2017}, like in ecological or food web networks, where it could be characterized WGs with similar topics by the corresponding motifs counts. Based on this approach, new methods can be developed to characterize and detect the presence of annoying users, spammers, bots, or fake profiles.

\section*{Acknowledgment} 
\small
The authors thank the Sao Paulo Research Foundation (FAPESP) grants 18/24260-5, 16/23698-1, 18/01722-3, 19/00157-3, 17/05831-9, 16/16291-2, and 15/50122-0; CNPq grant 313426/2018-0, and DFG-GRTK grant 1740/2 for the financial support. 

\bibliographystyle{IEEEtran}

\end{document}